\documentclass[aps,10pt,twocolumn,noshowpacs,prl]{revtex4-1}
\usepackage{hyperref}
\usepackage{amsmath}
\usepackage{graphicx}
\usepackage{epsfig}
\usepackage{latexsym,amssymb,bm}
\usepackage{epstopdf}
\usepackage{color}

\begin{document}

\title{Metasurfaces with maximum chirality empowered by bound states in the continuum}

\author{Maxim V. Gorkunov$^1$, Alexander A. Antonov$^1$, and Yuri S. Kivshar$^{2}$}
\affiliation{$^1$Shubnikov Institute of Crystallography of the Federal Scientific Research Centre ``Crystallography and Photonics'', Russian Academy of Science, Moscow 119333, Russia\\
$^2$Nonlinear Physics Centre, Australian National University, Canberra ACT 2601, Australia}

\begin{abstract}
We demonstrate that rotationally symmetric chiral metasurfaces can support arbitrarily sharp resonances with the maximum optical chirality determined by precise shaping of bound states in the continuum (BICs). Being uncoupled from one circular polarisation of light and resonantly coupled to its counterpart, a metasurface hosting the chiral BIC resonance exhibits a narrow peak in the circular dichroism spectrum. We propose a realization of such \textit{chiral BIC metasurfaces} based on pairs of dielectric bars and validate the concept of maximum chirality by numerical simulations.
\end{abstract}


\maketitle


Chirality refers to a global property of many systems which do not coincide with their mirror images. Photonic structures made of chiral elements exhibit chiroptical effects such as optical dichroism for left and right circularly polarized light, the property highly suitable for chiral nanophotonics~\cite{Schaeferling2017}.  However, strong chiroptical effects are challenging to achieve, and strong resonances of plasmonic structures~\cite{Oh2015} have been suggested for chiral mirrors~\cite{Plum2015,Hentschel2017}. Intrinsic chirality, i.e., chiroptical effects at normal incidence, are prohibited in planar structures, but can be observed in opaque metallic structures of complex shapes~\cite{Gorkunov2014} and dielectric layers facilitating chiral excitation of higher-order multipoles~\cite{Zhu2017}.

Recently, all-dielectric metasurfaces have been employed to achieve sharp optical resonances empowered by the physics of {\em bound states in the continuum} (BICs) when light at the resonance remains localized in the metasurface even though the state coexists with a continuum of electromagnetic waves~\cite{Hsu2016,Koshelev2019}.  In practice, BICs are realized with high but finite quality factors due to structural losses and imperfections, and they are usually termed “quasi BICs” (or q-BICs as termed below). The BIC-inspired resonances in the symmetry-broken all-dielectric metasurfaces are receiving attention for many applications \cite{Koshelev2018,filiz2019,Koshelev2019a}. To date, all metasurfaces supporting q-BIC resonances were associated with the transition between BICs and radiative continuum due to in-plane symmetry breaking. However, to achieve the chiral response, we need to manipulate the transition between BICs and leaky resonances utilizing out-of-plane perturbations, which allow controlling chiral properties.

\begin{figure}
	\centering\includegraphics[width=6cm]{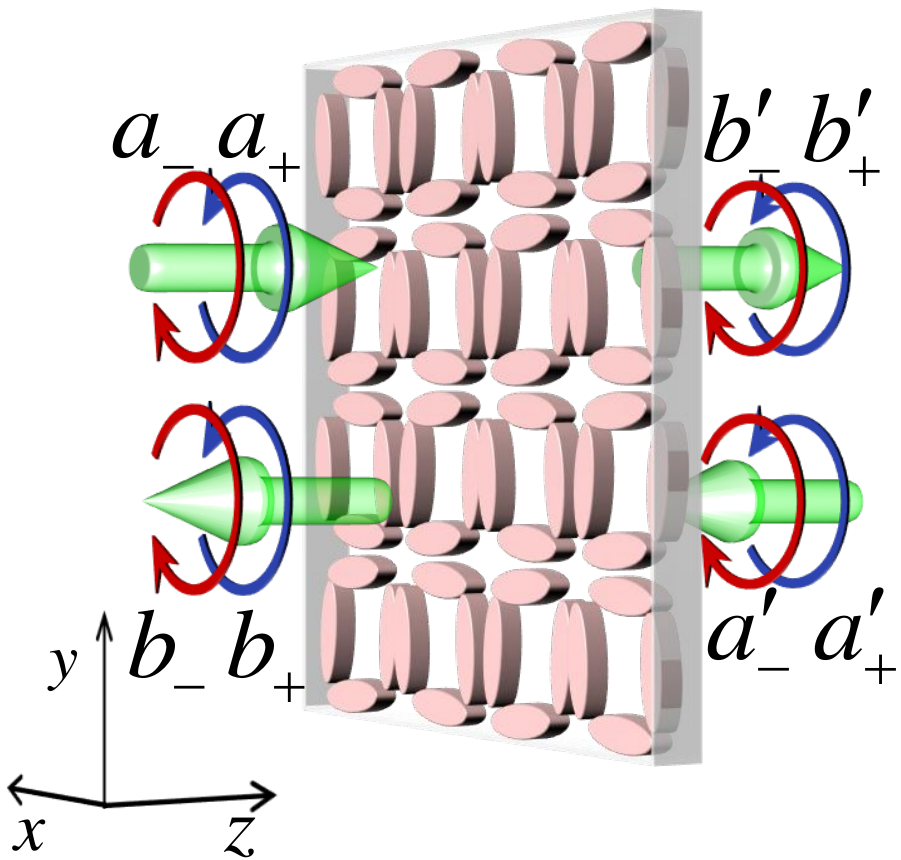} \caption{Sketch of the transmission-reflection problem for rotationally symmetric chiral metasurfaces described by the S-matrix equation~\eqref{SpmEq}.}\label{fig:scheme}
\end{figure}

In this Letter, we introduce the concept of highly transparent chiral metasurfaces by shaping BICs into q-BICs of maximum optical chirality delivering arbitrarily narrow peak of unit height in the circular dichroism spectrum. We propose a design based on dimers of dielectric bars and validate the maximum chirality of its q-BIC resonance by numerical modelling.

We consider a metasurface shown schematically in Fig.~\ref{fig:scheme}, located in the $xy$-plane with the $z$-axis being the $N$-th order rotational symmetry axis ($C_N$ symmetry group). For $N\ge3$, all polarization transformations of incoming waves incident normally from both sides are determined by the chirality. Assuming $e^{-i\omega t}$ time dependence of all fields, we consider waves of certain helicity polarized along the complex unit vectors:
\begin{equation}\label{epm}
{\bf e}_\pm=({\bf e}_x\mp i {\bf e}_y)/\sqrt{2}.
\end{equation}
For waves propagating in the positive $z$ direction, ${\bf e}_+$ and ${\bf e}_-$ correspond to the right circular polarization (RCP) and left circular polarization (LCP) respectively. For waves in the negative $z$ direction, the opposite is true.

Most generally, this transmission-reflection problem is described by an S-matrix equation  (see Supplemental Material \cite{Sup}) relating the outgoing wave amplitudes  $b_\pm$ and $b'_\pm$ with the incident wave amplitudes $a_\pm$ and $a'_\pm$ as:
\begin{equation}\label{SpmEq}
\left(
\begin{array}{c}
b_+ \\
b'_+ \\
b_- \\
b'_- \\
\end{array}
\right)=
\left(
\begin{array}{cccc}
r & t_L & 0 & 0\\
t_R & r' & 0 & 0\\
0 & 0 & r & t_R\\
0 & 0 & t_L & r'\\
\end{array}
\right)\left(
\begin{array}{c}
a_+ \\
a'_+ \\
a_- \\
a'_- \\
\end{array}
\right),
\end{equation}

Fundamental principle of reciprocity together with the rotational symmetry put substantial restrictions on the metasurface transmission and reflection \cite{Gorkunov2015, Kondratov2016}. In particular, there is no conversion of wave helicity (the property known also as duality~\cite{Fernandez-Corbaton2016}), the RCP and LCP transmission amplitudes, $t_R$ and $t_L$ respectively, are equal for both sides of incidence along with the key optical parameters, optical rotation (OR) and circular dichroism (CD):
\begin{equation} \label{CD_OR_def}
OR = \frac{1}{2}(\arg t_L - \arg t_R),\
CD = \frac{|t_{R}|^2 - |t_{L}|^2}{|t_{R}|^2 + |t_{L}|^2}.
\end{equation}
The reciprocity also determines that the reflection amplitudes $r$ and $r'$ are helicity independent, which directly relates the CD with the energy dissipation. Indeed, absorbed parts of the energy of incident RCP or LCP waves,
$A_{R,L}=1-|t_{R,L}|^2-|r|^2$,
determine that the transmittance difference
$|t_{R}|^2 - |t_{L}|^2=A_L-A_R$
arises due to a difference in dissipation.

The inherent connection of the CD with dissipation naturally determines a quantitative condition of maximum optical chirality: it is achieved when \textit{the metasurface is fully transparent for one circular polarization and totally absorbs its counterpart}.

To specify a feasible route to the maximum chirality, we employ the phenomenological coupled-mode theory (CMT) allowing expressing abstract scattering amplitudes in terms of physically meaningful parameters~\cite{Yoon2012,Alpeggiani2017}. Generalizing CMT for chiral metasurfaces with  plasmon~\cite{Kondratov2016} and dielectric~\cite{Gorkunov2018} resonances reproduced their observed strong chirality and clarified its origin.

In CMT, transmission and reflection are split into background and resonant channels with the latter determined by excitation and irradiation of certain eigenstates hosted by the structure. Note that the particular eigenstate normalization is not required \cite{Alpeggiani2017}.
CMT yields (see these and following relations derived in Supplemental Material~\cite{Sup}) the transmission amplitudes:
\begin{equation}\label{tLR}
t_R=\tau-\frac{m_+m'_-}{i(\omega-\omega_0)-\gamma_0},\
t_L=\tau-\frac{m'_+m_-}{i(\omega-\omega_0)-\gamma_0},
\end{equation}
where $\tau$ is the background transmission amplitude, the resonance frequency $\omega_0$ and damping $\gamma_0$ are helicity independent, and $m_\pm$ are the parameters of coupling of the eigenstates to the waves of corresponding helicity incident on one metasurface side, and $m'_\pm$ are those for the other side.
Accordingly, the optical chirality \eqref{CD_OR_def} is determined by the chirality of eigenstate coupling to the free-space continuum.

All losses contribute to the damping $\gamma_0=\gamma_d+\gamma_r$, where $\gamma_d$ is its dissipative part, and the radiative part is determined by the coupling:
\begin{equation}\label{gammar}
2\gamma_r=|m_+|^2+|m'_+|^2=|m_-|^2+|m'_-|^2.
\end{equation}
The difference of transmittances expressed as
\begin{equation}\label{tRtLn}
|t_R|^2-|t_L|^2=2\gamma_d\frac{|m_{-}|^2-|m_{+}|^2}{(\omega-\omega_0)^2+(\gamma_r+\gamma_d)^2},
\end{equation}
emphasizes the crucial role of dissipation for the CD.

Consider, for definiteness,  how to maximize the optical chirality by enhancing $|t_R|$ and suppressing $|t_L|$. Eq.~\eqref{tLR} suggests to achieve the ultimate value $|t_R|=1$ by setting $|\tau|=1$ and $m_+m'_-=0$. The latter condition requires uncoupling the eigenstate from waves of certain helicity on a particular metasurface side.

For example, we set $m_+=0$ and, according to \eqref{tRtLn},
\begin{equation}\label{tR2}
|t_L|^2=1-2\gamma_d\frac{|m_{-}|^2}{(\omega-\omega_0)^2+(\gamma_r+\gamma_d)^2}.
\end{equation}
with the minimum reached at the resonance, $\omega=\omega_0$:
\begin{equation}\label{tLCOMmin}
\min|t_L|^2=1-\frac{8\gamma_d|m_-|^2}{(|m_-|^2+|m'_-|^2+2\gamma_d)^2}.
\end{equation}
where the radiative damping \eqref{gammar} is substituted. The ultimate limit $\min|t_L|^2=0$ is achieved only if simultaneously $m'_-=0$ and $|m_-|^2=2\gamma_d$.

Therefore, the maximum chirality requires $m_+=m'_-=0$, i.e., eigenstates selectively decoupled from the free-space continuum. The second condition reduces by Eq.~\eqref{gammar} to
\begin{equation}\label{critcoupl}
\gamma_r=\gamma_d
\end{equation}
which is a classical \textit{critical coupling regime}~\cite{Haus1984} of a resonator receiving from the continuum exactly the amount of energy it is capable to dissipate.

To summarize, CMT unambiguously points out that for the maximum optical chirality a metasurface has to:
\begin{itemize}
	\item host eigenstates selectively coupled to circularly polarized waves: e.g. the ``+'' state uncoupled from the RCP waves incident on one side and the ``--'' state uncoupled from the LCP waves incident on the other side;
	\item fully absorb the light of opposite circular polarizations in the critical coupling regime.
\end{itemize}
Note that the phenomenological approach requires a metasurface performing identically from its both sides, e.g., being coupled with $m_+=0$ and $|m_-|^2=2\gamma_d$ to the waves on one side and with $m'_-=0$ and $|m'_+|^2=2\gamma_d$ on the other side. This can be automatically ensured by restricting to flipping symmetric designs.

Achieving maximum optical chirality requires precise control of the coupling of eigenstates to the free-space continuum.
In the following, we present a step-by-step design of such eigenstates starting from fully uncoupled BICs and carefully enabling their selective coupling by symmetry breaking perturbations.

\begin{figure}
	\centering\includegraphics[width=0.9\columnwidth]{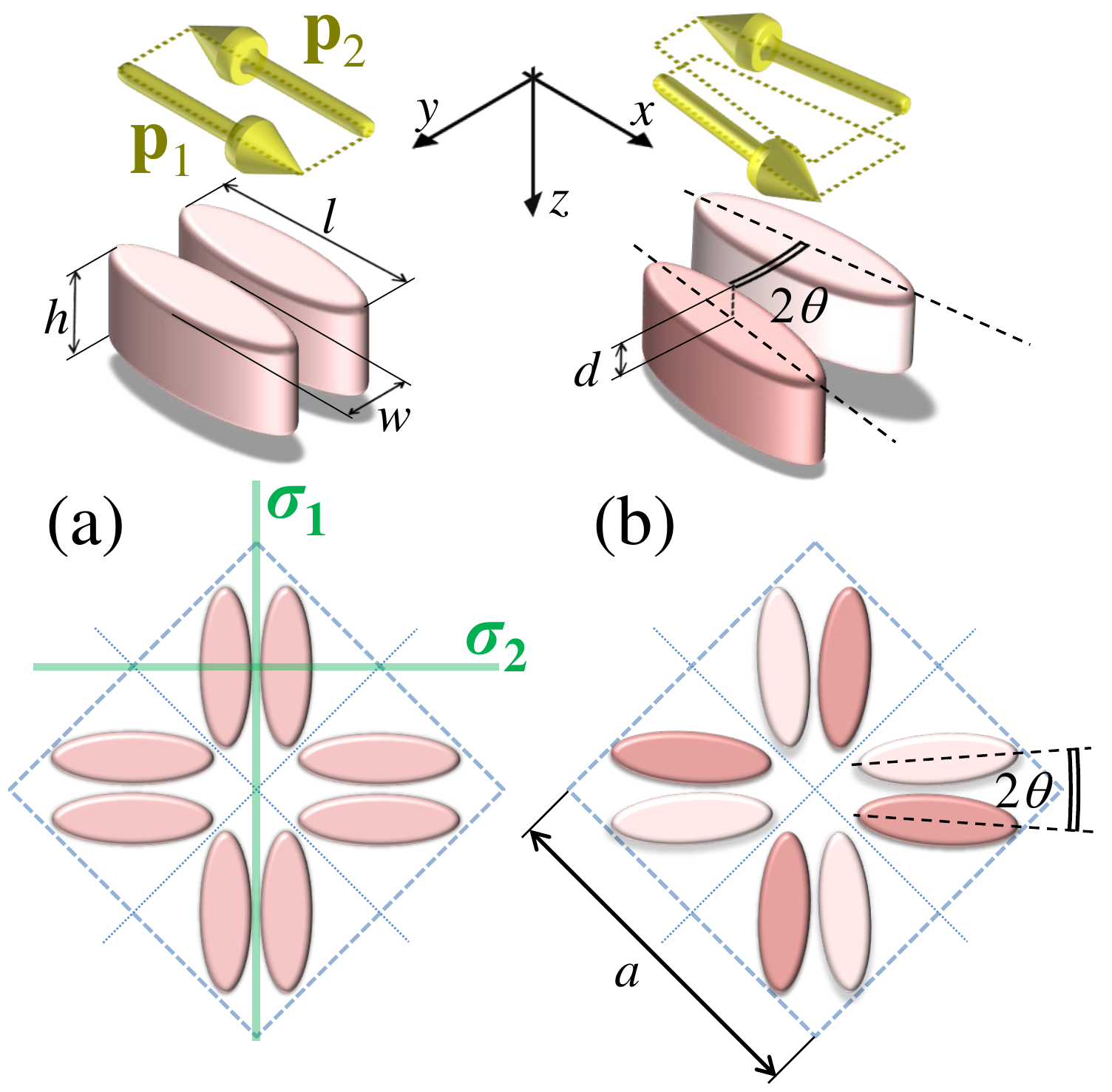}
	\caption{Shaping q-BIC chirality by symmetry breaking: (a) a dimer of parallel bars and their lattice hosting BIC resonances; (b) a dimer of bars vertically offset by $d$ and rotated by $\theta$, and a unit cell of their lattice hosting chiral q-BIC resonances. All bars are identical and the colours indicate location on different levels. Two types of mirror symmetry planes are shown by $\sigma$-lines in (a). The lattice constant $a$ is shown in (b). Relative orientation and offset of electric dipole moments characterising BIC and q-BIC eigenstates are shown on the top for each dimer type.} \label{fig:Vdimers}
\end{figure}

Consider a planar lattice of pairs of parallel dielectric bars shown in Fig.~\ref{fig:Vdimers}(a). The lattice is reflection symmetric with respect to three types of mirror planes: those indicated as $\sigma_1$ and $\sigma_2$, and the plane of drawing. All dimers are in perfectly symmetric situations as the vertical $C_2$ axes (intersection of planes $\sigma_1$ and $\sigma_2$) pierce exactly through their centres. All resonant eigenmodes transform according to the irreducible representations of the $C_2$ group and, among those, there is a symmetric $A$-representation. The electric resonance of this symmetry is described by a pair of antiparallel dipole moments ${\bf p}_1=-{\bf p}_2$ shown on the top of Fig.~\ref{fig:Vdimers}(a).

To estimate the coupling parameter of a dimer eigenstate to a vertically incident plane wave polarized along a unit vector $\bf e$ and having a wavevector ${\bf k}$ along the $z$-axis,
one can integrate the incident wave field with the eigenstate field or, equivalently, with its current density~\cite{Alpeggiani2017,Koshelev2018}:
\begin{equation}\label{mgen}
m_{\bf e}\propto \int_{V_1,V_2} d{\bf r}\ ({\bf j}({\bf r})\cdot {\bf e}) \ e^{ikz},
\end{equation}
where $V_{1,2}$ are the volumes of dielectric bars.
For each volume, the integral yields the dipole moment of the corresponding bar, and the antiparallel dipole eigenstate is uncoupled of all incident polarizations as:
\begin{equation}\label{mpar}
m_{\bf e}\propto{\bf p}_1\cdot{\bf e}+{\bf p}_2\cdot{\bf e}=0.
\end{equation}
This is a perfect BIC with respect to all normally incident waves.

Introducing weak symmetry breaking transforms BIC into q-BIC. As has been studied in much detail~\cite{Koshelev2018}, diverging the bars by an in-plane angle $\theta$ eliminates the mirror symmetry plane $\sigma_2$ and the rotational symmetry $C_2$. The coupling parameters estimated as integrals of the eigenstate currents with the wave fields polarized along ${\bf e}_\pm$:
\begin{equation}\label{mtwist}
m_\pm\propto{\bf p}_1\cdot{\bf e}_\pm+{\bf p}_2\cdot{\bf e}_\pm=i\sqrt{2}p\sin\theta.
\end{equation}
are helicity independent, as the structure retains certain mirror symmetries.

An out-of-plane symmetry breaking can be introduced by a small vertical offset $d$ of bars within each dimer eliminating the mirror plane $\sigma_1$ together with the $C_2$ axis. The corresponding coupling parameters:
\begin{equation}\label{moffset}
m_\pm\propto{\bf p}_1\cdot{\bf e}_\pm+{\bf p}_2\cdot{\bf e}_\pm e^{ikd}=i\sqrt{2} pe^{ikd/2}\sin{{kd}/{2}}.
\end{equation}
are also achiral, as certain mirror symmetries remain.

Combining the offset by $d$ with the rotation by $\theta$, as illustrated in Fig.~\ref{fig:Vdimers}(b), breaks all mirror symmetries. Packing the dimers in the depicted square lattice ensures the $C_4$ rotational and flipping symmetries. The coupling parameters estimated as
$m_\pm\propto{\bf p}_1\cdot{\bf e}_\pm+{\bf p}_2\cdot{\bf e}_\pm e^{ikd}
=i\sqrt{2} pe^{ikd/2}\sin\left({kd}/{2}\mp\theta\right)$,
elucidate the rise of optical chirality.
For the flipping symmetry, the coupling to the  waves incident on the other side is the same up to a helicity interchange:
$m'_\pm\propto i\sqrt{2} pe^{ikd/2}\sin\left({kd}/{2}\pm\theta\right)$.

Remarkably, maximizing the q-BIC chirality is possible by a simple adjustment of the offset and rotation as
\begin{equation}\label{MainVd}
\theta=kd/2,
\end{equation}
which ensures $m_+=m'_-=0$. Under this condition, the remaining coupling parameters
\begin{equation}\label{mrest}
m_-=m'_+\propto i\sqrt{2}pe^{i\theta}\sin(2\theta)
\end{equation}
are controlled by $\theta$ and allow continuous tuning to match the dissipation in the critical coupling regime \eqref{critcoupl}.


\begin{figure*}
	\centering\includegraphics[width=\textwidth]{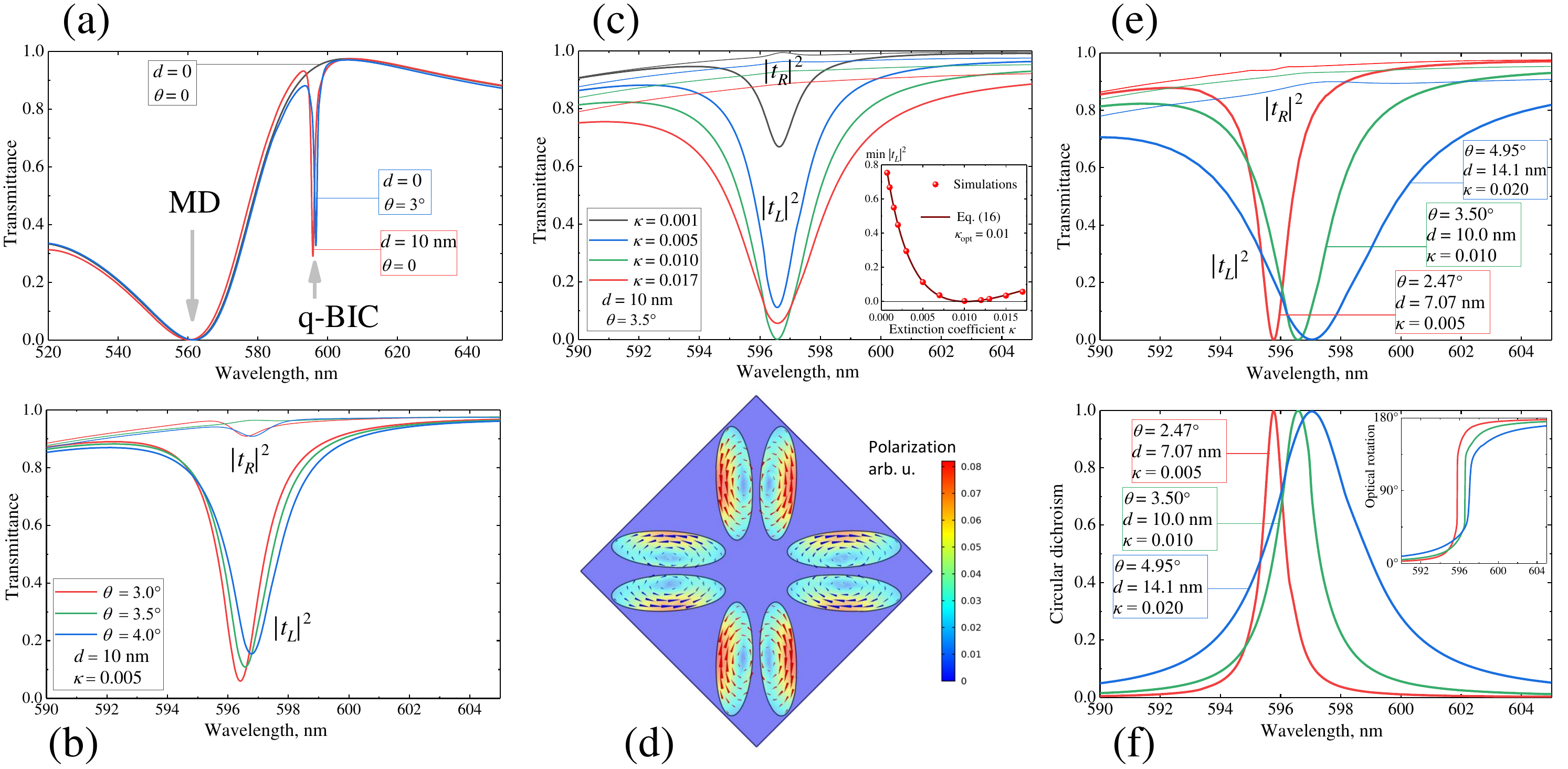}
	\caption{Simulated transformation of BIC into maximum chiral q-BIC. Transmittance spectra of achiral structures (a) comprised of parallel bar dimers, dimers rotated by $\theta$ and dimers offset by $d$, exhibit strong MD resonance and sharp q-BIC resonance controlled by rotation and offset. RCP and LCP transmittance spectra (b) of chiral structures with equal offset and different rotation angles. RCP and LCP transmittance spectra (c) of structures with q-BIC uncoupled from RCP waves for different extinction coefficient $\kappa$ with the inset comparing simulated minimum transmittance with that predicted by CMT Eq.~\eqref{tLminkappa}. Polarisation distribution (d) in maximum chiral q-BIC for $\kappa=0.01$, $\theta=3.5^\circ$, and  $d=10$~nm excited by linearly polarized light of $\lambda=596.5$~nm: distribution of real (red cones) and imaginary (blue cones) polarisation vector components are juxtaposed with the colourmap of average polarization. Spectra of transmittances (e), CD (f), and OR (inset in (f)) of structures hosting maximum chiral q-BICs with symmetry breaking parameters and losses following the scaling rule $\theta^2 \propto d^2 \propto \kappa$.} \label{fig:Vdimernumerics}
\end{figure*}

To validate the symmetry-based analysis, we numerically study a particular structure using Comsol Multiphysics. To keep relevance with available materials (silicon and germanium), we model bars consisting of dielectric having complex refractive index $n=4+i\kappa$ with small extinction coefficient $\kappa$. For simplicity, the background refractive index is set to unity. The dimensions of elliptical bars are $l=220$~nm, $w=70$~nm and $h=100$~nm (see Fig.~\ref{fig:Vdimers}). The gap between parallel bars is set to 15 nm and the square lattice constant $a=480$~nm excludes diffraction in the considered wavelength range.

First, we simulate the transmission of normally incident light by achiral lattices of parallel bar dimers, dimers diverged by an in-plane rotation by $\theta=3^\circ$, and dimers with an out-of-plane offset $d=10$ nm. In all three cases, the transmission is polarization insensitive.
As seen in Fig.~\ref{fig:Vdimernumerics}(a), the transmittance spectra stay generally very close exhibiting a broad minimum at about the 560~nm wavelength associated with the magnetic-dipole (MD) resonance. Close to the 596~nm wavelength, either of the two weak symmetry perturbations give rise to much sharper q-BIC resonances fully inline with the estimates (\ref{mpar}-\ref{moffset}).

To determine the parameters necessary for a q-BIC uncoupled from RCP waves, we use Eq.~\eqref{MainVd} to estimate that the offset $d=10$~nm at a 596~nm wavelength requires the angle $\theta=3.0^\circ$. Indeed, as shown in Fig.~\ref{fig:Vdimernumerics}(b), for this angle, the resonance of the RCP transmittance is weak. However, the ideal uncoupling occurs for $\theta=3.5^\circ$. Further increasing the angle to $\theta=4.0^\circ$ restores the resonance.
We conclude that Eq.~\eqref{MainVd} specifies geometries very close to optimal, though a small mismatch arises due to a slight misalignment of the bar dipole moment and shape.

Knowing the perfect combination of $\theta$ and $d$, we proceed to verify the main CMT conclusions. For a proper chiral q-BIC state isolated from RCP waves by $m_+=m'_-=0$, the minimum LCP transmittance at the resonance is given by Eq.~\eqref{tLCOMmin} with $m'_-=0$. The remaining coupling parameter $|m_-|^2$ is determined by the bar refractive index, dimensions and packing, while weak dissipation affects it negligibly. The damping, on the contrary, is determined by the dissipation as $\gamma_d\propto\kappa$.

If a certain value $\kappa=\kappa_{\rm opt}$ corresponds to the critical coupling regime, Eq.~\eqref{tLCOMmin} can be expressed as:
\begin{equation}\label{tLminkappa}
\min|t_L|^2=\left(\frac{\kappa_{\rm opt}-\kappa}{\kappa_{\rm opt}+\kappa}\right)^2,
\end{equation}
Accordingly, we keep all other parameters fixed and simulate the transmission of  structures with different $\kappa$, as illustrated in Fig.~\ref{fig:Vdimernumerics}(c). Decreasing $\kappa$ drastically elevates the $|t_L|^2$ minimum as the losses are critical for the optical chirality. Increasing $\kappa$ far above 0.01 also negatively affects the chirality, elevates the $|t_L|^2$ minimum and suppresses $|t_R|^2$. Plotting $\min|t_L|^2$ values as a function of $\kappa$ and fitting them by the simple dependence \eqref{tLminkappa} (see the inset in Fig.~\ref{fig:Vdimernumerics}(c)), validates the CMT conclusions and precisely determines $\kappa_{\rm opt}=0.01$. The inner structure of the chiral q-BIC state is illustrated in Fig.~\ref{fig:Vdimernumerics}(d) by the polarisation distribution within a unit cell of the structure illuminated by linearly polarized wave. Only the LCP part interacts with the chiral q-BIC and excites the polarisation currents of the ${\bf e}_-$ helicity: anticlockwise rotation by $\pi/2$ is equivalent to multiplication by $i$.

Finally, combining Eqs.~\eqref{mrest} and \eqref{MainVd} reveals a general rule: the maximum chirality is established with the three small parameters scaling as  $\theta^2 \propto d^2 \propto \kappa$. To verify this, we simulate the transmission of structures with doubled and halved $\kappa$ and with $\theta$ and $d$ varied accordingly by a factor of $\sqrt{2}$. As shown in Fig.~\ref{fig:Vdimernumerics}(e), the structures indeed host q-BICs with maximal chirality, their CD spectra in Fig.~\ref{fig:Vdimernumerics}(f) possess resonances of unit height accompanied by typical OR kinks shown in the inset.


As the Si refractive index at $\lambda=595$~nm is $n_{\rm Si}= 3.948+0.021i$ \cite{Jellison1992}, the broader resonance in Figs.~\ref{fig:Vdimernumerics}(f,e) describes a practically available realization. Lower extinction coefficients of 0.01 and 0.005 correspond to Si at $\sim700$~nm and $\sim800$~nm, respectively. Sharper CD resonances available in the near IR range require fabrication technique supporting precise nanometer-scale offsets.

In conclusion, we have developed the concept of chiral BIC metasurfaces transmitting one circular polarisation and resonantly blocking the opposite polarization. Our design strategy is applicable for maximizing the optical chirality of
other types of metasurfaces operating in the visible and near IR spectral ranges.

\begin{acknowledgments}
The work of MVG and AAA	is supported by the Ministry of Science and Higher Education of the Russian Federation within the State assignment of FSRC ``Crystallography and Photonics'' RAS. YSK acknowledges the support from the Strategic Fund of the Australian National University. The authors are grateful to Alexey Kondratov for his assistance with Comsol Multiphysics modelling.
\end{acknowledgments}

%

\end{document}


\title{Supplemental Material: \\Metasurfaces with maximum chirality empowered by bound states in the continuum}

\author{Maxim V. Gorkunov$^1$, Alexander A. Antonov$^1$, and Yuri S. Kivshar$^{2}$}
\affiliation{$^1$Shubnikov Institute of Crystallography of the Federal Scientific Research Centre ``Crystallography and Photonics'', Russian Academy of Sciences, Moscow 119333, Russia\\
	$^2$Nonlinear Physics Centre, Australian National University, Canberra ACT 2601, Australia}

\begin{abstract}
In the Supplemental Material we (i) analyze the S-matrix symmetry and derive Eq.~(2) of the main text, and (ii) provide a rigorous formulation of CMT yielding Eqs.~(6--9) of the main text.
\end{abstract}


\maketitle

\setcounter{equation}{0}
\setcounter{figure}{0}
\setcounter{table}{0}
\setcounter{page}{1}
\makeatletter
\renewcommand{\theequation}{S\arabic{equation}}
\renewcommand{\thefigure}{S\arabic{figure}}
\renewcommand{\bibnumfmt}[1]{[S#1]}
\renewcommand{\citenumfont}[1]{S#1}

\section{S-matrix of rotationally symmetric chiral metasurface}\label{sec:Smatr}

\begin{figure}
	\centering\includegraphics[width=7cm]{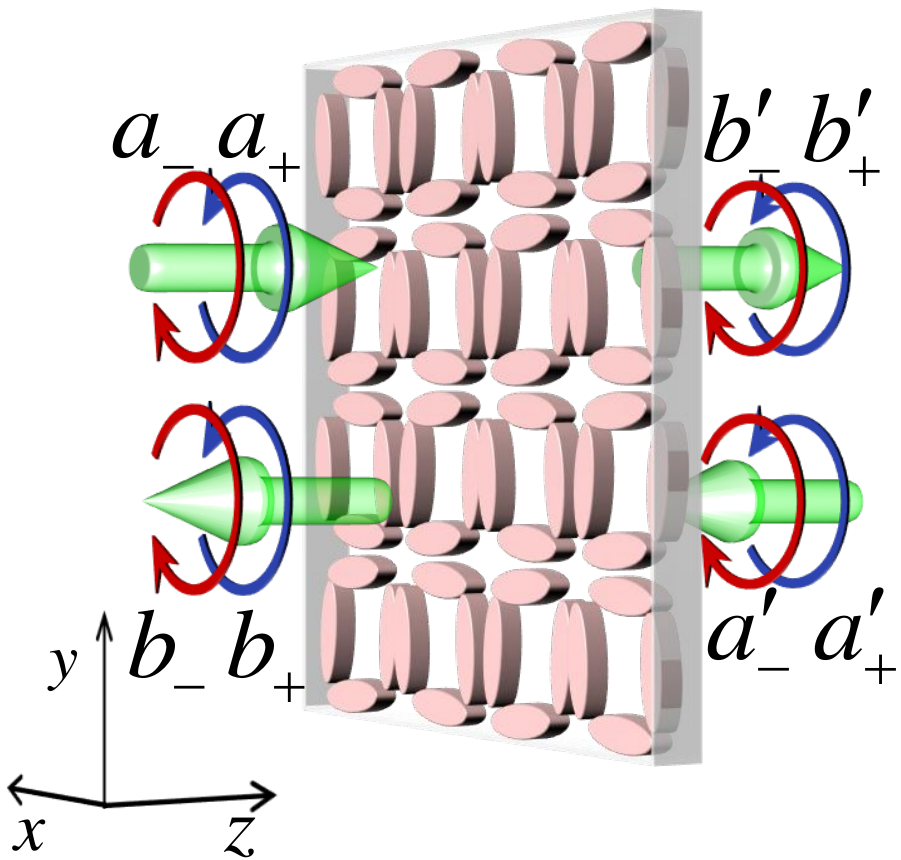} \caption{Sketch of the transmission-reflection problem for a rotationally symmetric metasurface  described by Eq.~\eqref{SpmEq}.}\label{fig:scheme}
\end{figure}

Consider a light transmitting metasurface possessing a rotational symmetry axis of the $N$-th order ($C_N$-symmetry) with incoming transversal waves incident normally on its both sides. As we show, the main advantage of rotationally symmetric metasurfaces with $N\ge3$ is that they affect the light polarization only due to their intrinsic chirality.

To reveal the most general features of light transmission and reflection following from the fundamental principles of symmetry and reciprocity, we start with the S-matrix in the basis of linear polarizations. For the transversal waves propagating along the $z$-axis, the S-matrix equation reads as
\begin{equation}\label{Sxygen}
{\bf b}={\bf S}{\bf a},\ {\rm or\ expressed\ by\ components:}\
\left(
\begin{array}{c}
b_x \\
b'_x \\
b_y \\
b'_y \\
\end{array}
\right)=
\left(
\begin{array}{cccc}
S_{11} & S_{12} & S_{13} & S_{14}\\
S_{12} & S_{22} & S_{23} & S_{24}\\
S_{13} & S_{23}  & S_{33}  & S_{34}\\
S_{14}  & S_{24} & S_{34} & S_{44}\\
\end{array}
\right)\left(
\begin{array}{c}
a_x \\
a'_x \\
a_y \\
a'_y \\
\end{array}
\right)
\end{equation}
where $a_i$ and $a'_i$ are the components of the amplitudes of waves incident on different sides, and $b_i$ and $b'_i$ are those of the waves outgoing from different sides. The S-matrix is symmetric due to Lorentz reciprocity.

$N$-th order rotational symmetry implies the S-matrix invariance with respect to a rotation about the $z$-axis by an angle $\phi=2\pi/N$. Using corresponding coordinate transform matrix
\begin{equation}\label{Tphi}
{\bf T}_\phi=\left(
\begin{array}{cccc}
\cos\phi & 0 & \sin\phi & 0\\
0 & \cos\phi & 0 & \sin\phi\\
-\sin\phi & 0 & \cos\phi  & 0 \\
0  & -\sin\phi & 0 & \cos\phi\\
\end{array}
\right)
\end{equation}
and requiring that ${\bf S}^{(xy)}={\bf T}_\phi {\bf S}^{(xy)} {\bf T}_\phi^T$ one obtains $S_{33}=S_{11}$, $S_{44}=S_{22}$, $S_{34}=S_{12}$, $S_{23}=-S_{14}$, $S_{13}=S_{24}=0$, which reduces the S-matrix to:
\begin{equation}\label{Sxy}
{\bf S}^{(xy)}=\left(
\begin{array}{cccc}
S_{11} & S_{12} & 0 & S_{14}\\
S_{12} & S_{22} & -S_{14} & 0\\
0 & -S_{14}  & S_{11}  & S_{12} \\
S_{14}  & 0 & S_{12} & S_{22}\\
\end{array}
\right).
\end{equation}

Transformation from the Cartesian coordinate basis ${\bf e}_{x,y}$ to the helical basis ${\bf e}_\pm=({\bf e}_x\mp i {\bf e}_y)/\sqrt{2}$ is performed by the matrix
\begin{equation}\label{Tpm}
{\bf T}_c=\frac{1}{\sqrt{2}}\left(
\begin{array}{cccc}
1 & 0 & -i & 0\\
0 & 1 & 0 & -i\\
1 & 0  & i  & 0 \\
0  & 1 & 0 & i\\
\end{array}
\right),
\end{equation}
and the S-matrix in the ``$\pm$'' basis obtained as ${\bf S}={\bf T}_c{\bf S}^{(xy)}{\bf T}^\dag_c$ reads:
\begin{equation}\label{Spm}
{\bf S}=\\
\left(
\begin{array}{cccc}
S_{11} & S_{12}+iS_{14} & 0 & 0\\
S_{12}-iS_{14} & S_{22} & 0 & 0\\
0 & 0  & S_{11}  & S_{12}-iS_{14} \\
0 & 0 & S_{12}+iS_{14} & S_{22}\\
\end{array}
\right).
\end{equation}

The analogue of Eq.~\eqref{Sxygen} in the helical basis reads as:
\begin{equation}\label{SpmEq}
\left(
\begin{array}{c}
b_+ \\
b'_+ \\
b_- \\
b'_- \\
\end{array}
\right)=
\left(
\begin{array}{cccc}
r & t_L & 0 & 0\\
t_R & r' & 0 & 0\\
0 & 0 & r & t_R\\
0 & 0 & t_L & r'\\
\end{array}
\right)\left(
\begin{array}{c}
a_+ \\
a'_+ \\
a_- \\
a'_- \\
\end{array}
\right),
\end{equation}
which is introduced as Eq.~(2) and allows identifying the matrix elements in Eq.~\eqref{Spm} with LCP and RCP transmission amplitudes, $t_L$ and $t_R$, and the amplitudes of reflection from different metasurface sides, $r$ and $r'$.
The block diagonal form of the S-matrix in Eq.~\eqref{SpmEq}
determines independence of the transmission-reflection problems for the $``+"$ and $``-"$ helicities.

Therefore, the symmetry and reciprocity require that the transmission amplitudes of waves of opposite helicity incident on different metasurface sides (same circular polarization) are equal, and that the reflection amplitudes are helicity independent, i.e., achiral.

\section{Chiral coupled-mode theory}\label{sec:CMT}

Metasurface optical response is dominated by its resonances. Coupled-mode theory (CMT) provides very convenient description of such systems and allows deriving a particular form of the S-matrix expressed by a few phenomenological parameters of clear meaning \cite{Yoon2012, Kondratov2016}.
The usual key assumption of the CMT approach is the weakness of dissipation losses. Then, at the first step, one considers a lossless metasurface, implies the requirements of symmetry, reciprocity and reversibility, and establishes many useful relations between the model parameters. Next, weak dissipation losses are introduced providing certain corrections and, in particular, giving rise to the optical chirality.

Metasurface eigenstates can be formally obtained as solutions of source-free Maxwell's equations:
\begin{equation}\label{eigenstates}
\nabla\times\nabla\times{\bf E}^{(j)}({\bf r})-\varepsilon({\bf r})\frac{\omega_j^2}{c^2} {\bf E}^{(j)}({\bf r})=0,
\end{equation}
where the states are enumerated by the index $j$ and outgoing wave conditions at $z\rightarrow\pm\infty$ are imposed on ${\bf E}^{(j)}$. The metasurface structure is described by inhomogeneous permittivity $\varepsilon({\bf r})$ The inherent dissipation and radiative energy losses determine small negative imaginary part of the eigenfrequencies $\omega_j$.
The metasurface rotational symmetry determines the eigenstate ${\bf E}^{(j)}({\bf r})$ symmetry. The connection is universally described by the group theory apparatus involving appropriate irreducible point group representations \cite{Hopkins2015}.

Consider the consequences of the $C_4$ metasurface rotational symmetry implying its invariance upon rotation by $\pi/2$ about the $z$-axis. The rotation transforms the coordinates as $x'= y$, $y'= -x$ and $z'= z$ and the metasurface symmetry means that  $\varepsilon(x,y,z)=\varepsilon(y,-x,z)$. Upon such rotation, the eigenstate field components interchange and may acquire phase factors:
\begin{eqnarray}
E_{x}^{(j)}(x,y,z)=\exp(i\pi j/2)\ E_y^{(j)}(y,-x,z),\\
E_{y}^{(j)}(x,y,z)=-\exp(i\pi j/2)\ E_x^{(j)}(y,-x,z),\\
E_{z}^{(j)}(x,y,z)=\exp(i\pi j/2)\ E_z^{(j)}(y,-x,z).
\end{eqnarray}
where only the numbers $j=-1,0,1,2$ are meaningful and ensure that performing the rotation 4 times yields the initial state.

As eigenstates of an open resonator, ${\bf E}^{(j)}({\bf r})$ posses the far field asymptotics of transversal plane waves outgoing in the $\pm z$ directions, i.e., having the nonzero components $E_{x}^{(j)}(z)$ and $E_{y}^{(j)}(z)$ depending only on the $z$-coordinate. Not all numbers $j$ are compatible with this. For $j=0$, one must simultaneously satisfy $E_{x}^{(0)}(z)=E_{y}^{(0)}(z)$ and $E_{x}^{(0)}(z)=- E_{y}^{(0)}(z)$. For $j=2$, a pair of other incompatible equalities, $E_{x}^{(2)}(z)=-E_{y}^{(2)}(z)$ and $E_{x}^{(2)}(z)=E_{y}^{(2)}(z)$, must hold true. Therefore, the eigenmodes with $j=0,2$ are true BIC states topologically isolated from the continuum.

The eigenmodes with $j=\pm 1$ asymptotically transform into plane waves obeying the relations $E_{x}^{(\pm 1)}(z)=\pm iE_{y}^{(\pm 1)}(z)$, i.e., polarized with a certain helicity. By the symmetry, each of the two eigenmodes is coupled to one of the two independent transmission-reflection problems described by Eq.~\eqref{SpmEq}. As has been shown in Ref.~\cite{Hopkins2015},  Lorentz reciprocity ensures that their eigenfrequencies are equal, i.e., $\omega_{\pm 1}=\omega_0-i \gamma_0$.

To describe the interaction of metasurface with electromagnetic radiation oscillating with a real frequency $\omega$, one characterizes  the eigenstates by slow-varying amplitudes $P_\pm$, and their particular normalization in a phenomenological CMT is of minor importance \cite{Alpeggiani2017}. Introducing for compactness the vector of these amplitudes ${\bf P} = (P_-,P_+)^T$ we write the equation describing how the incident waves excite the eigenstates as:
\begin{equation}\label{COM1}
\frac{d\textbf{P}}{dt }=[i(\omega-\omega_0)-\gamma_0]{\bf P}+{\bf M} {\bf a},
\end{equation}
where the matrix
\begin{equation}\label{M}
{\bf M}=\left(\begin{array}{cccc}
m_+& m'_+ & 0 & 0 \\
0 & 0 & m_- & m'_-
\end{array}\right)
\end{equation}
describes the coupling of the incident waves with the eigenmodes,
and the vector of the incident wave amplitudes $\bf a$ is the same as in the r.h.s. of Eq.~\eqref{SpmEq}.

Similarly, the irradiation of the outgoing waves is described by another CMT equation:
\begin{equation}\label{COM2}
{\bf b}={\bf N}{\bf P}+{\bf C}{\bf a},
\end{equation}
where the vector of outgoing wave amplitudes $\bf b$ is the same as in the l.h.s. of Eq.~\eqref{SpmEq}, while the matrix
\begin{equation}\label{N}
{\bf N}=\left(\begin{array}{cc}
m_-& 0\\
m'_- & 0 \\
0  & m_+\\
0 & m'_+
\end{array}\right)
\end{equation}
stands for the irradiation of circularly polarized waves by the corresponding resonances and is related to the matrix \eqref{M} by Lorentz reciprocity.
The S-matrix of direct transmission is supposed to be achiral and free of dissipation:
\begin{equation}\label{Cmatr}
{\bf C}=\left(
\begin{array}{cccc}
\rho & \tau & 0 & 0\\
\tau & \rho & 0 & 0\\
0 & 0 & \rho & \tau\\
0 & 0 & \tau & \rho\\
\end{array}
\right),
\end{equation}
where $|\tau|^2+|\rho|^2=1$. The stationary state of Eqs.~\eqref{COM1} and \eqref{COM2} determines that the S-matrix from Eq.~\eqref{SpmEq} reads as:
\begin{equation}\label{SCOM}
{\bf S}={\bf C}- \frac{{\bf N}{\bf M}}{i(\omega-\omega_0)-\gamma_0}.
\end{equation}

\subsection{Coupled-mode theory without dissipation}

Totally neglecting the eigenstate energy dissipation losses one may require Eqs.~\eqref{COM1} and \eqref{COM2} to be symmetric with respect to the time reversal.
Time reversibility of the background S-matrix requires its unitarity: ${\bf \hat C}^\dag={\bf \hat C}^{-1}$. Note that for a symmetric matrix, the Hermitian transpose is equal to the complex conjugate: ${\bf \hat C}^\dag={\bf \hat C}^*$.
If a set $\bf a$, $\bf b$, and ${\bf P}$ satisfies Eqs.~\eqref{COM1} and \eqref{COM2}, so should do their time-reversed counterparts $\tilde{\bf a}$, $\tilde{\bf b}$, and $\tilde{\bf P}$ expressed as:
\begin{equation}\label{abpinv}
{\tilde{\bf a}} = {\bf T}{\bf b}^*, \ {\tilde{\bf b}} = {\bf T}{\bf a}^*,\  {\tilde{\bf P}} = {\bf T}_P{\bf P}^*,
\end{equation}
with the help of matrices ${\bf T}$ and ${\bf T}_P$ producing the helicity interchange $+\leftrightarrow -$ :
\begin{equation}
{\bf T}=\left(
\begin{array}{cccc}
0& 0 & 1 & 0\\
0 & 0 & 0 & 1\\
1& 0 & 0 & 0\\
0& 1 & 0 & 0\\
\end{array}
\right),\
{\bf T}_P=\left(
\begin{array}{cc}
0 & 1 \\
1& 0\\
\end{array}
\right).
\end{equation}

Substituting the amplitudes \eqref{abpinv} into Eq.~\eqref{COM2} and multiplying by the matrix $\bf T$ from the left side yields:
\begin{equation}\label{COM2inv}
{\bf a}^*={\bf T}{\bf N}^*{\bf T}_P{\bf P}^*+{\bf b}^*,
\end{equation}
where we have used that ${\bf T}^2=1$, ${\bf T}{\bf C}{\bf T}={\bf C}$ for the matrix given by Eq.~\eqref{Cmatr}.
Noticing that ${\bf T}{\bf N}^*{\bf T}_P={\bf M}^\dag$ for the matrices given by Eq.~\eqref{M} and Eq.~\eqref{N} and taking the complex conjugate, we obtain
\begin{equation}\label{COM2inv2}
{\bf b}=-{\bf C}{\bf M}^\dag{\bf P}+{\bf C}{\bf a}.
\end{equation}
Comparing it with Eq.~\eqref{COM2} one identifies an important condition:
\begin{equation}\label{CNM}
{\bf C}{\bf {M}}^\dag=-{\bf N}.
\end{equation}

The set of time-reverse amplitudes \eqref{abpinv} also obeys Eq.~\eqref{COM1} with the inverse time derivative sign. Multiplying it with ${\bf T}_P$ from the left and taking the complex conjugate yields
\begin{equation}\label{COM1inv}
-\frac{d{\bf P}}{dt}=[-i(\omega-\omega_0)-\gamma_0]{\bf P}+{\bf N}^\dag{\bf b}.
\end{equation}
where the property ${\bf T}_P{\bf M}^*{\bf T}={\bf N}^\dag$ has been used. Now one can substitute the amplitudes $\bf b$ as defined by Eq.~\eqref{COM2} and obtain:
\begin{equation}\label{COM1inv2}
\frac{d{\bf P}}{dt}=[i(\omega-\omega_0)+\gamma_0]{\bf P}-{\bf N}^\dag{\bf N}{\bf P}-{\bf N}^\dag{\bf C}{\bf a}.
\end{equation}
Expressing \eqref{CNM} as ${\bf {N}}^\dag{\bf C}=-{\bf M}$ and comparing with Eq.~\eqref{COM1} one concludes that the time reversal symmetry requires that
\begin{equation}\label{NN}
2\gamma_0=2\gamma_r={\bf N}^\dag{\bf N}=|m_+|^2+|m'_+|^2=|m_-|^2+|m'_-|^2,
\end{equation}
In the absence of dissipation, the damping constant $\gamma_0$ equals the radiative damping constant $\gamma_r$ directly determined by the eigenstate coupling to the free-space continuum.

\subsection{Coupled-mode theory with a weak dissipation}

Now we introduce weak dissipation losses and consider the S-matrix \eqref{SCOM} when the eigenstates experience both dissipative and radiative damping so that their eigenfrequencies are $\omega_0-i\gamma_0=\omega_0-i(\gamma_r+\gamma_d)$.
Other model parameters are supposed to remain unaffected by the weak dissipation.
Then using the relations \eqref{CNM} and \eqref{NN} one can evaluate
\begin{multline}\label{Sunitn}
{\bf S}^\dag{\bf S}=
=\left[{\bf C}^\dag-\frac{{\bf M}^\dag{\bf N}^\dag}{-i(\omega-\omega_0)-\gamma_0}\right]
\left[{\bf C}-\frac{{\bf N}{\bf M}}{i(\omega-\omega_0)-\gamma_0}\right]
={\bf 1}+\left[\frac{1}{-i(\omega-\omega_0)-\gamma_0}+\frac{1}{i(\omega-\omega_0)-\gamma_0}\right]{\bf M}^\dag{\bf M}\\+
\frac{{\bf M}^\dag{\bf N}^\dag{\bf N}{\bf M}}{|i(\omega-\omega_0)-\gamma_0|^{2}}
= {\bf 1}-\frac{2\gamma_0{\bf M}^\dag{\bf M}}{(\omega_0-\omega)^2+{\gamma_0^2}}+
\frac{2\gamma_r{\bf M}^\dag{\bf M}}{(\omega_0-\omega)^2+{\gamma_0^2}}
={\bf 1}-\frac{2\gamma_d{\bf M}^\dag{\bf M}}{(\omega_0-\omega)^2+\gamma_0^2},
\end{multline}
which shows that the S-matrix loses its unitarity in the presence of $\gamma_d$. This breaks the energy conservation together with the reversibility of the system.

According to the S-matrix equation \eqref{SpmEq}, the main diagonal of the difference  ${\bf  1}-{\bf  S}^\dag{\bf S}$ contains the absorption rates for different circular polarizations and sides of incidence. From Eq.~\eqref{Sunitn} we can express them as:
\begin{equation}\label{AR}
A_R=1-|r|^2-|t_R|^2=\frac{2\gamma_d|m_{+}|^2}{(\omega_0-\omega)^2+\gamma_0^2},
\end{equation}
\begin{equation}\label{AL}
A_L=1-|r|^2-|t_L|^2=\frac{2\gamma_d|m_{-}|^2}{(\omega_0-\omega)^2+\gamma_0^2},
\end{equation}
\begin{equation}
A'_R=1-|r'|^2-|t_R|^2=\frac{2\gamma_d|m'_{-}|^2}{(\omega_0-\omega)^2+\gamma_0^2},
\end{equation}
\begin{equation}
A'_L=1-|r'|^2-|t_L|^2=\frac{2\gamma_d|m'_{+}|^2}{(\omega_0-\omega)^2+\gamma_0^2}.
\end{equation}
i.e., as all exhibiting the same Lorentzian frequency dispersion.

The transmission amplitudes can be explicitly  expressed as components of the S-matrix \eqref{SCOM}:
\begin{equation}\label{StR}
t_R=\tau-\frac{m_+m'_-}{i(\omega-\omega_0)-\gamma_0}
\end{equation}
\begin{equation}\label{StL}
t_L=\tau-\frac{m'_+m_-}{i(\omega-\omega_0)-\gamma_0}
\end{equation}
Note that each of these values appears independently as two elements of the S-matrix in Eq.~\eqref{SpmEq}. The equality of those elements follows from Lorentz reciprocity and it is fulfilled as we have thoroughly imposed all reciprocity requirements in CMT equations.

A relation very helpful for understanding the limits of optical chirality can be obtained by taking the difference of Eqs.~\eqref{AL} and \eqref{AR}:
\begin{equation}\label{StRtLn}
|t_R|^2-|t_L|^2=2\gamma_d\frac{|m_{-}|^2-|m_{+}|^2}{(\omega-\omega_0)^2+(\gamma_r+\gamma_d)^2}.
\end{equation}
It emphasizes the crucial role of dissipative damping for chirality and shows that the frequency dispersion of CD thas also a clear Lorentzian shape. The maximum CD value is reached at $\omega=\omega_0$.